\documentclass[prl,twocolumn,superscriptaddress,floatfix,amsmath,amssymb]{revtex4-1}
\usepackage[utf8]{inputenc}
\usepackage{graphicx}
\usepackage{hyperref}
\usepackage{color}
\usepackage{bm}
\usepackage[normalem]{ulem}

\usepackage{tikz}
\usetikzlibrary{arrows,graphs,positioning}
\usetikzlibrary{decorations.pathreplacing}

\newcommand{\integral}[3]{\int_{#2}^{#3} \!\! \mathrm{d} #1 \,}
\newcommand{\difffrac}[2]{\frac{\mathrm{d} #1}{\mathrm{d} #2}}
\newcommand{\partialfrac}[2]{\frac{\partial #1}{\partial #2}}

\newcommand{\ket}[1]{\left| {#1} \right\rangle}

\newcommand{\proj}[2]{\left| {#1} \right\rangle\!\left\langle {#2} \right|}
\renewcommand{\vec}{\bm}

\newcommand{\intHd}[1]{H_{I,\,#1}}
\newcommand{\intH}{H_{I}}
\newcommand{\stripintHd}[1]{h_{I,\,#1}}
\newcommand{\stripintH}{h_{I}}

\newcommand{\ii}{\mathrm{i}}
\newcommand{\tr}{\operatorname{Tr}}

\newcommand{\exptval}[1]{\left\langle {#1} \right\rangle}
\newcommand{\comm}[2]{\left[{#1},{#2}\right]}

\renewcommand{\Re}{\mathrm{Re}} 
\renewcommand{\Im}{\mathrm{Im}} 

\newcommand{\itsection}[1]{ {\it {#1}.-}}

\begin{document}

\title{Information transmission without energy exchange}

\author{Robert H. Jonsson}
\affiliation{Department of Applied Mathematics, University of Waterloo, Waterloo, Ontario, N2L 3G1, Canada}
\author{Eduardo Mart\'{i}n-Mart\'{i}nez}
\affiliation{Department of Applied Mathematics, University of Waterloo, Waterloo, Ontario, N2L 3G1, Canada}
\affiliation{Institute for Quantum Computing, University of Waterloo, Waterloo, Ontario, N2L 3G1, Canada}
\affiliation{Perimeter Institute for Theoretical Physics, 31 Caroline St N, Waterloo, Ontario, N2L 2Y5, Canada}
\author{Achim Kempf}
\affiliation{Department of Applied Mathematics, University of Waterloo, Waterloo, Ontario, N2L 3G1, Canada}
\affiliation{Institute for Quantum Computing, University of Waterloo, Waterloo, Ontario, N2L 3G1, Canada}
\affiliation{Perimeter Institute for Theoretical Physics, 31 Caroline St N, Waterloo, Ontario, N2L 2Y5, Canada}

\begin{abstract}
We show that it is possible to use a massless field in the vacuum to communicate in such a way that the signal travels  arbitrarily slower than the speed of light and such that no energy is transmitted from the sender to the receiver. Instead, the receiver has to supply a signal-dependent amount of work to switch his detector on and off. Because of that, this kind of communication without energy exchange may be called ``Quantum Collect Calling''. This type of  communication is related to Casimir-like interactions and it is made possible by dimension ---and curvature--- dependent subtleties of Huygens' principle. 
\end{abstract}

\maketitle

Communication  by means of massless quantum fields is normally discussed in terms of the exchange of field quanta. Typically, a sender emits photons ---which carry energy  along null-geodesics--- that are then absorbed by a receiver.  The receiver has to intercept the beam of light, or he misses the signal.  In this letter we introduce a method of communication via  massless quantum fields which breaks with each point of the just mentioned intuitive understanding:  1) No emission or absorption of real quanta is necessary;  2) the information flow need not be carried by an energy flow; and  3) the information flow can be  arbitrarily slower than light, in spite of the field being massless. 

The fact that waves of massless fields propagate at exactly the speed of light is merely  a peculiar feature of $n+1$D Minkowski spacetime, for odd  $n\geq 3$. When there is generic curvature, or when the spatial dimension is even, then waves of massless fields propagate both on as well as  \emph{inside} the lightcone  \cite{McLenaghan,czapor,poisson}. Here, we translate this classical wave phenomenon into quantum field theory (QFT) and analyze the implications for quantum  communication.

Surprisingly, when applied to QFT, we find that the resulting signal transmission effects appear already at the leading order in  perturbation theory. These are not emission and absorption processes of real photons, which only appear at subleading order. Indeed, we find that slower-than-light quantum signals do not require the transmission of energy from the sender to the receiver. Instead ---much as for a collect call--- the receiver has to provide  energy  for the detection of the signal.

\itsection{The strong Huygens' principle in QFT}
The Green's functions of the massless Klein-Gordon field in 3+1D Minkowski space have support only on the lightcone. Hence, any disturbances of the field propagate strictly along null geodesics. This classical phenomenon is called the strong Huygens' principle \cite{McLenaghan,czapor}. 

In QFT,  communication via the field is possible if and only if the commutator between the field at two events does not vanish \cite{AK-Cliche1,AK-Cliche2,ourfirstpaper,fulling,ooguri1986spectrum}. The commutator is given by the Green's function $\comm{\phi(\vec{x},t_1)}{\phi(\vec{y},t_2)}=\ii   G(\vec{x},t_1,\vec{y},t_2)\openone$, where $G=G_{adv}-G_{ret}$; thus the commutator always vanishes outside the lightcone.
The strong Huygens' principle therefore manifests itself in QFT through the commutator of the field: In  3+1D  Minkowski space, the commutator is only  supported on the lightcone.
In 1+1D and 2+1D, the commutator is non-vanishing inside the lightcone. Namely, in 1+1D it takes the constant value,
$\comm{\phi(x,t_1)}{\phi(y,t_2)}=\ii/2,$ 
when $(y,t_2)$ is in the future lightcone of $(x,t_1)$. In 2+1 dimensions, the commutator takes these values inside the lightcone: 
\begin{align}\label{commt}
\comm{\phi(\vec{x},t_1)}{\phi(\vec{y},t_2)} =\frac{\ii}{2\pi} \frac{\mathrm{sgn}(t_2-t_1)}{\sqrt{(t_1-t_2)^2 - |\vec x - \vec y |^2}}
\end{align}
Hence,  communication between timelike separated points, which is impossible in 3+1D flat spacetime, is possible in 1+1 and $2n$+1 dimensions, and in generic, even 3+1D,  curved spacetimes.

\itsection{Setup}
 We study  communication between two  Unruh-DeWitt (UDW) detectors  \cite{DeWitt,AK-Cliche1,AK-Cliche2}. This model captures fundamental features of light-matter interactions when there is no exchange of orbital angular momentum \cite{Wavepackets,Alvaro,ScullyBook}.  We consider two pointlike detectors, referred to as Alice and Bob, both at rest at some fixed distance $L$ from each other.  The total interaction Hamiltonian $H_I=H_{I,A}+H_{I,B}$, is the sum of the two detector-field interaction Hamiltonians. In the interaction picture:  
\begin{equation}\label{eq:intHd}
\intHd{d}(t) = \lambda_d \, \eta_d (t) \, m_d(t) \,  \phi(\vec{x}_d,t)
\end{equation}
Here, $d\in\{A,B\}$, and $\lambda_d$ is the coupling strength. $\eta_d(t)$ controls the switching, $m_d(t)=\proj{e}{g}e^{\ii \Omega_d t}+\proj{g}{e}e^{-\ii \Omega_d t}$ is the detector's monopole moment, and $\phi(\vec{x}_d,t)$ is the field operator at the detector's location.  The detectors' free Hamiltonian is $H_d=\Omega_d\proj{e_d}{e_d}$ with energy gap $\Omega_d$.

\begin{figure}[tb]
\centering
\begin{tikzpicture}
\def\c{1.5} 

\def\a{2};
\def\b{4};
\def\TA{0.8};
\def\TBon{1.8};
\def\TBoff{2.6};

\def\xmin{-0.2};
\def\xmax{7};
\def\tmax{2.8};
    \draw[->] (\xmin,0)--(\xmax,0) node [below] {$x$};
    \draw[->] (0,-0.2)--(0,{\tmax*1.05}) node[left] {$t$};
    \draw[thick,|-|] (\a,-0.015)-- node [left] {Alice} (\a,{\TA+0.015});
    \draw[thick,|-|] (\b,{\TBon-0.015}) -- node [left] {Bob} (\b,{\TBoff+0.015});
\draw[decorate,decoration={brace, amplitude=3pt},xshift=0pt,yshift=-2pt]  (\b,0) -- (\a,0)   node [below,midway,yshift=-1pt] {$L$};

\draw[shift={(0,\TA)}] (0pt,0pt) -- (-2pt,0pt) node [left] {$T_A$};
\draw[dashed] (0,\TA) -- (\a,\TA);

\draw[shift={(0,\TBon)}] (0pt,0pt) -- (-2pt,0pt) node [left] {$T_1$};
\draw[dashed] (0,\TBon) -- (\b,\TBon);

\draw[shift={(0,\TBoff)}] (0pt,0pt) -- (-2pt,0pt) node [left] {$T_2$};
\draw[dashed] (0,\TBoff) -- (\b,\TBoff);
\draw[dotted] (\a,0) -- ({\a+\tmax*\c},\tmax);
\draw[dotted] (\a,0) -- (\xmin,{(\a-\xmin)/\c});

\draw[dotted] (\a,\TA) -- ({\a+(\tmax-\TA)*\c},\tmax);
\draw[dotted] (\a,\TA) -- (\xmin,{\TA+(\a-\xmin)/\c});

\end{tikzpicture}
\caption{Space-time diagram of the setup. The dotted lines indicate the first and the last light rays emanating from Alice. The distance $L$ between the detectors determines whether they are spacelike, null or timelike separated.}
\label{fig:stdiagram}
\end{figure}
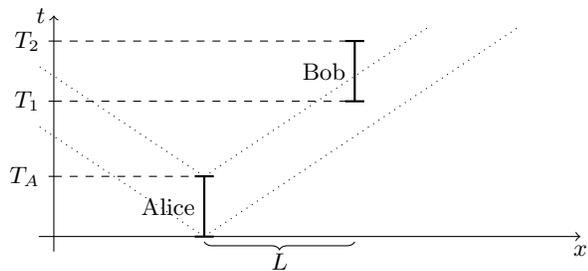

We calculate the time evolution of the system using perturbation theory \cite{ourfirstpaper}.
 Alice, the sender, couples to the field first. We assume that Alice switches off her detector before Bob, the receiver, switches his detector on. Fig. \ref{fig:stdiagram} depicts the setup: We assume  that we have
$\text{supp } \eta_A \subseteq [0,T_A]$ and $\text{supp } \eta_B \subseteq [T_1, T_2]$ with $T_A<T_1$, for the support of the switching functions.  As initial state we assume a  product of  an arbitrary state of the field $\rho$ and pure states for Alice's and Bob's detectors 
\begin{align}
\ket{A_0}=\!\alpha_A \ket{e_A}\!+\! \beta_A \ket{g_A},\;
\ket{B_0}=\!\alpha_B \ket{e_B}\!+\! \beta_B \ket{g_B}.
\end{align}

\itsection{Timelike signalling} We split Bob's excitation probability  $P_e(t)$ into three contributions: Bob's excitation probability $\left|\alpha_B\right|^2$ due to his initial state, Bob's excitation probability $R(t)$ due to noise, and the {\it signalling contribution} $S(t)$,
\begin{align}\label{eq:exprob}
P_e(t)=\left| \alpha_B \right|^2 + R(t) + S(t).
\end{align}
Here, the noise-caused probability, $R(t)$, for Bob's detector to be found excited has two sources: one source of noise is the switching that couples Bob to the field. The second source of noise in $R(t)$ is the quantum fluctuations of the field that Bob picks up once Bob is coupled to the field. $R(t)$ is independent of Alice, i.e., it  contains no dependence on $\lambda_A$ \cite{ourfirstpaper}.

The signalling contributions, $S(t)$, determine how Bob's excitation probability is modulated by Alice's interaction with the field. Expanding $S(t)\sim \lambda_A \lambda_B S_2  +\mathcal{O}(\lambda^3)$, its leading order contribution is
\begin{align}
\label{eq:signallingPe}
S_2 &= 4  \integral{t_2}{T_1}t \eta_B(t_2) \integral{t_1}0{T_A} \eta_A(t_1)  \Re\left(\alpha_A^* \beta_A e^{\ii \Omega_A t_1}\right)\nonumber\\ 
& \times  \Re \Big( \alpha_B^* \beta_B e^{\ii \Omega_B t_2} \tr \big( \rho\comm{\phi(\vec{x}_A,t_1)}{\phi(\vec{x}_B,t_2)} \big) \Big).
\end{align}
Notice that, in \eqref{eq:signallingPe}, the trace, i.e., the expectation of the commutator, is independent of the state of the field. Therefore, while $S$ is, of course, dependent on the initial state of the field, its leading order contribution, $S_2$, is independent of the state of the field.

Note that for $S_2$ to be the leading order contribution to $S(t)$ we require that neither of the detectors starts out in an energy eigenstate: $\alpha_{d}^*\beta_{d}\neq 0,\ d\in\{A,B\}$.

The signal, carried by $S(t)$, competes with the quantum noise $R(t)$,  and  the question arises whether there is a threshold  level of noise above which communication is no longer possible.
We will now show that,  within the perturbative regime, for any noise level, a finite channel capacity can be achieved straightforwardly.  Correspondingly, we will also show  that Bob has a better than 50\% success probability of guessing correctly an equally distributed random bit sent by Alice.

Among the many ways in which Alice and Bob can encode and decode a message, the following simple protocol already provides a lower bound to the channel capacity: Let  Alice encode `1' or `0' by either or not coupling her detector to the field.   Bob afterwards temporarily couples his detector to the field, then measures the energy of his detector. Depending on whether Alice's coupling  increases or decreases Bob's excitation probability, Bob  interprets finding the excited (or ground) state as his logical `1' (or  `0'), or vice versa.

\begin{figure}
\centering
\begin{tikzpicture}

\def\rshift{1.75}
\def\rrshift{4.5}
\def\ushift{0.7}
\def\sshift{0.35}

\coordinate (A) at (0,0); 
\node[font=\bf] (A1) at (\rshift,\ushift) {Alice `1'}; 
\node[font=\bf] (A0) at (\rshift, -\ushift) {Alice `0'};

\node[font=\bf] (A1B1) at (\rrshift,\ushift-\sshift) {Bob `1'};
\node[font=\bf] (A1B0) at (\rrshift,\ushift+\sshift) {Bob `0'};

\node[font=\bf] (A0B1) at (\rrshift,-\ushift-\sshift) {Bob `1'};
\node[font=\bf] (A0B0) at (\rrshift,-\ushift+\sshift) {Bob `0'};

\draw[->] (A) to node[above,sloped] {$1/2$} (A1) ;
\draw[->] (A) to node[below,sloped] {$1/2$} (A0);

\draw[->] (A1) to node[below,sloped] {p} (A1B1);
\draw[->] (A1) to node[above,sloped] {1-p} (A1B0);
\draw[->] (A0) to node[below,sloped] {$q$} (A0B1);
\draw[->] (A0) to node[above,sloped] {$1-q$}(A0B0);

\draw[decorate,decoration={brace, amplitude=3pt},xshift=0pt,yshift=-2pt]  (A1B1.east) -- (A0B0.east) node [right,midway,xshift=+2pt, align=left] {success probability\\ $\quad\frac12 p +\frac12 (1-q)$}; 

\end{tikzpicture}
\caption{ Tree diagram for Bob guessing a random bit sent by Alice. The  probability for  a correct guess is given by the sum of the probabilities for the two correct outcomes out of the four possible outcomes.
Since  in our setup $p=q+|S(t)|$,
 his success probability is $\frac12 + \frac12 (p-q) = \frac12+|S(t)|>\frac12$.}
\label{fig:channeldiag}
\end{figure}
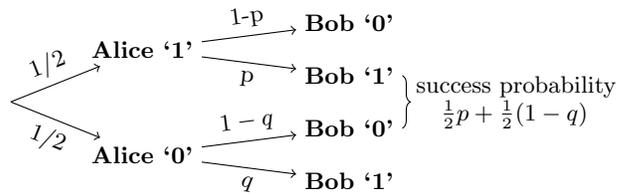

Now, when Alice encodes a `0' (through her inaction), Bob has finite probabilities ($q$ and $1-q$) to find either `1' or `0', due to quantum noise and his detector's initial probability distribution. If Alice chooses to send a `1' (by temporarily coupling her detector to the field), then this changes Bob's outcome probabilities to $p$ and $1-p$. In our case, $q=\left|\alpha_B\right|^2+R(t)$ and $p=q+\left| S(t)\right|$.  Hence, the  channel  allows Bob to improve his  probability at guessing Alice's bit to $\frac12 + \left| S(t) \right|$, see Fig. \ref{fig:channeldiag}.

We can also use this system to set up a binary asymmetric channel between Alice and Bob. These channels have  the following Shannon capacity \cite{silverman}
\begin{align}\label{capacity}
C= \frac{-q\, h(p) + p\, h(q)}{q-p} + \log_2 \left( 1+2^{\frac{h(p)-h(q)}{q-p}}\right),
\end{align}
where $h(x)=-x \log_2(x)-(1-x) \log_2(1-x)$.

Crucially, the capacity is non-zero whenever $p\neq q$, as is the case for our channel whenever the signalling contribution $S(t)$ is non-vanishing. This means that the  channel between Alice and Bob possesses a finite capacity that is never completely drowned out by the noise. In fact, using the leading order signalling contribution $S_2$, the channel capacity can be expanded as
\begin{align}\label{expan}
C\sim \lambda_A^2 \lambda_B^2 \frac2{\ln{2}} \left(\frac{S_2}{4 \left|\alpha_B\right| \left| \beta_B \right|}\right)^2 +\mathcal{O}\left(\lambda^6\right) 
\end{align}
which, at leading order, is independent of the field's initial state and shows a positive  contribution to the channel capacity if $S_2\neq 0$.

As expected, the capacity \eqref{capacity} is not symmetric in $\lambda_A, \lambda_B$  ($p$ depends only on $\lambda_B$ and $q$ depends on  $\lambda_A$ and $\lambda_B$), while the small coupling expansion in \eqref{expan} happens to be symmetric in $\lambda_A$ and $\lambda_B$. Interestingly, as $C\sim \lambda_A^2\lambda_B^2$, in the weak coupling regime, any increase of Alice's and Bob's couplings increases the capacity.

Note the appearance of the field commutator in \eqref{eq:signallingPe}  which shows that signalling between timelike separated UDW detectors is possible whenever the strong Huygens' principle does not hold: if the commutator between two points does not vanish, it is possible to couple two detectors around those points to the field such that \eqref{eq:signallingPe} does not vanish and thus signalling is possible.

Remarkably, the signalling contribution \eqref{eq:signallingPe} does not contain an exchange of real energy-carrying quanta. Indeed, energy-exchange processes (the sender's detector gets deexcited, emitting energy to the field, which later excites the receiver) have amplitudes  of order $\mathcal{O}(\lambda_A \lambda_B)$. Hence, they only contribute to  $P_e(t)$ at ${O}(\lambda^4)$ (see \cite{ourfirstpaper}).

Notice that, in 1+1D where the commutator is constant, arbitrarily many spacelike separated receivers  could be arranged in the future lightcone of Alice, such that all receive the same signalling contribution  to $P_e(t)$. Since the receivers are spacelike separated they cannot influence each other. This   shows that while the signalling contribution to $\exptval{H_B}$ is caused by Alice, it cannot be energetically sourced by her or it would violate energy conservation. We will see that the energetic cost of the excitation of Bob's detector is balanced by the work required to switch his detector.

\itsection{Energy Budget of Switching a Detector}
It is known that a single detector coupled to the vacuum can be excited by switching it on and off. The more sudden the switching, the larger the excitation probability \cite{Jormantes}.  In the following we compute how much work, on average,  is required to switch the detector.

When a single detector interacts with the field, the system's Hamiltonian  is given by $H(t)=H_f+H_d+\intH(t)$, the sum of the free field, free detector and the interaction Hamiltonian from \eqref{eq:intHd}.
Since we restrict ourselves to detectors at rest, the Hamiltonian is explicitly time dependent only through  $\eta(t)$, hence 
\begin{align}\label{eq:dHdt}
\difffrac{}t \exptval{H(t)} = \exptval{\partialfrac{\intH(t)}t}=  
\partialfrac{\eta(t)}t \exptval{\stripintH (t)},
\end{align}
where  $\stripintH(t)=\lambda m(t) \phi(\vec{x},t) $. 
Therefore, when the switching function is zero, the detector and the field are decoupled and each of the energy expectation values, $\exptval{H_f}$ and $\exptval{H_d}$, are conserved individually. For intervals during which $\eta(t)$ is constant, but non-vanishing, only its sum is conserved:
\begin{align}\label{eq:energybalance}
\exptval{H(t)}=\exptval{H_f}+\exptval{H_d}+\exptval{\intH(t)}=\text{const}.
\end{align}

Eq. \eqref{eq:dHdt} shows that the  total energy expectation  only changes when the coupling strength between field and detector is altered. 
Integrating \eqref{eq:dHdt}  yields the amount $\Delta\langle H\rangle$ by which a switching process changes the total energy expectation value. 
 This energy difference corresponds to the average work that a switching process requires.

For sharp switching functions, the time derivatives are given by $\delta$-functions. Hence,  $\Delta\langle H\rangle$ { is} given by $\langle \stripintH \rangle$. When the detector is switched on, the total energy expectation value rises by $\Delta\langle H\rangle=\langle{\stripintH(t=0)}\rangle$.
This is the average work it takes  to ``pull''  the detector's switch. When the detector is switched off, $\langle H \rangle$ drops by $\Delta\langle H\rangle=-\langle\stripintH (t=T)\rangle$.
Note that $\langle\stripintH(t)\rangle$ can be positive or negative, so in general a switching process can  either  cost or yield energy.
It is straightforward to show that the energy cost $\Delta\langle H\rangle$ for smooth switching functions converges to the cost of a sharp switching, $\langle\stripintH(t)\rangle$, as the switching functions approach sharp step functions.

If the  state before the switching is a product state of field and detector, the average energy cost of a sharp detector switch is given by the product $\exptval{\stripintH}=\lambda\exptval{\phi}\exptval{m}$.
Hence, it costs no energy to sharply switch a detector when the expectation value of the monopole operator vanishes, or  when  the field is in a Fock state. 
Therefore, in particular, when the field starts in the vacuum,  the  excitations that remain after the detector  is switched off, are entirely accounted for by the work it takes to decouple the detector from the field.

Notice also, that the fluctuations of the energy cost of  a sudden switching in the vacuum ($\propto\lambda^2 \langle m^2 \phi^2\rangle$) are unbounded. This is an artifact of pointlike detectors. This kind of divergences is well-known, see e.g. \cite{Jormantes,Wavepackets,Satz2006,Satz2007,space}. The spurious divergences dissappear when using smooth switching functions or spatially smeared detectors.

\itsection{Energy budget of signalling} 
The analysis of the previous section carries over to  multiple detectors. For simplicity, we will in the following sections consider that the field starts out in the vacuum state, i.e., $\rho=\proj{0}{0}$. The perturbative contributions of the time-evolved state fulfill \eqref{eq:dHdt} and \eqref{eq:energybalance} order by order in the coupling constants.

Among the leading order terms, the single detector contributions, $\mathcal{O}(\lambda_A^2)$ and $\mathcal{O}(\lambda_B^2)$, and the signalling contributions, $\mathcal{O}(\lambda_A \lambda_B)$, will each be balanced separately.  Thus, to understand the energy source for the timelike signals discussed earlier, we focus on the signalling contributions to the Hamiltonians' expectations. 

For times $t\geq T_1$, Alice remains uncoupled from the field. So only $\intHd{B}$ is relevant for the interaction Hamiltonian. Also, the signalling contributions to the detector Hamiltonian, $H_A$, cancel out  as expected since, otherwise, Bob would have means to signal into the past.

Assuming sharp switching (and $\rho=\proj00$), the relevant  signalling contributions, for $T_1\leq t\leq T_2$, are
\begin{align}\label{eq:bobintHinterac}
&\exptval{\stripintHd{B}(t)}  \stackrel{\mathcal{O}(\lambda_A \lambda_B)}{=} 4\ii \lambda_A \lambda_B \Re\left(\alpha_B^* \beta_B e^{\ii \Omega_B t} \right) \nonumber\\
&\quad\times \integral{t_1}{0}{T_A} \Re\left(\alpha_A^* \beta_A e^{\ii \Omega_A t_1}\right) \comm{\phi(\vec{x}_A,t_1)}{\phi(\vec{x}_B,t)}
\end{align}
for the interaction Hamiltonian, whereas for $\exptval{H_B (t)}=\Omega_B P_e(t)$ the signalling contribution is given by \eqref{eq:signallingPe}.
The signalling contribution to the field Hamiltonian is
\begin{align}\label{eq:fieldHaminteract}
&\!\exptval{H_f}\!(t)\!\! \stackrel{\mathcal{O}(\lambda_A \lambda_B)}{=} \!\!  4 \lambda_A \lambda_B \!\! \int_{T_1}^t \!\!\! \mathrm{d}{t_2} \!\! \int_0^{T_A}\!\!\!\!\!\!\!\mathrm{d}t_1 \, \Re \left( \alpha_A^* \beta_A e^{\ii \Omega_A t_1}\right) \\
&\times\Re\left( \alpha_B^* \beta_B e^{\ii \Omega_B t_2} \right)\! \int\!\!\frac{\text{d}^n\vec{k}}{(2\pi)^n}\Re\left(  e^{\ii \left(k(t_1 - t_2) - \vec{k}\cdot(\vec{x}_A-\vec{x}_B)\right)} \right). \nonumber
\end{align}

The contributions to $\langle\stripintHd{B}\rangle$ and to $\langle H_B\rangle$  
are similar in that they both contain the commutator of the field.
The contributions to $\exptval{H_f}$ contain the integral kernel of a different distribution on spacetime. In 1+1D and 3+1D Minkowski space this distribution has support only on the lightcone. In 2+1 dimensions it has support on the lightcone and inside the future lightcone.

  Hence,  in  3+1D Minkowski space the signalling contributions vanish for timelike separated detectors, because neither the commutator nor the distribution in \eqref{eq:fieldHaminteract} have timelike support.  Crucially, however, in the presence of curvature, the commutator  has support in the time-like region, although it decays with distance \cite{poisson}.

Therefore, understanding the phenomena in lower-dimensional flat space may provide insight into the phenomena expected in curved 3+1D spacetimes: In 2+1D the commutator \eqref{commt} is proportional to the inverse geodesic distance inside the lightcone, a situation qualitatively similar to that of curved 3+1D spacetimes. 
  Notice also that in these cases, where the commutator decays into the lightcone, the channel naturally resets itself after each use. This means that the channel can easily be reused at some finite frequency. Remarkably, in 1+1D Minkowski space only   $\langle H_B\rangle$ and $\langle\stripintHd{B}\rangle$ receive signalling contributions for timelike separated detectors, whereas $\exptval{H_f}$ remains unchanged. Thus the 1+1D setting provides a clean testbed to study the relation between detector excitation and energy cost of switching the detector. The study of lower-dimensional cases may also be interesting considering that analog setups can implement massless fields in 1+1D and 2+1D, e.g., in circuit QED \cite{Wallraff,Solano,PastFutPRL} or graphene \cite{Vozmediano}.

\itsection{1+1 Dimensions} 
In 1+1 dimensions, the field energy is not affected by signalling contributions between two timelike separated detectors. Hence, any change in the average detector energy $\langle H_B\rangle$ is balanced by the interaction Hamiltonian $\langle\stripintHd{B}\rangle$ only: Through the interaction with the field, Alice changes the field state away from the vacuum. The imprint Alice leaves in the field determines the signalling contribution \eqref{eq:bobintHinterac} to $\langle\stripintHd{B} (t)\rangle$ at all later times  $t>T_A$. Hence,  Alice influences the average energy cost it takes Bob to switch his detector on or off.

\begin{figure}[tb]
\includegraphics[width=0.44\textwidth]{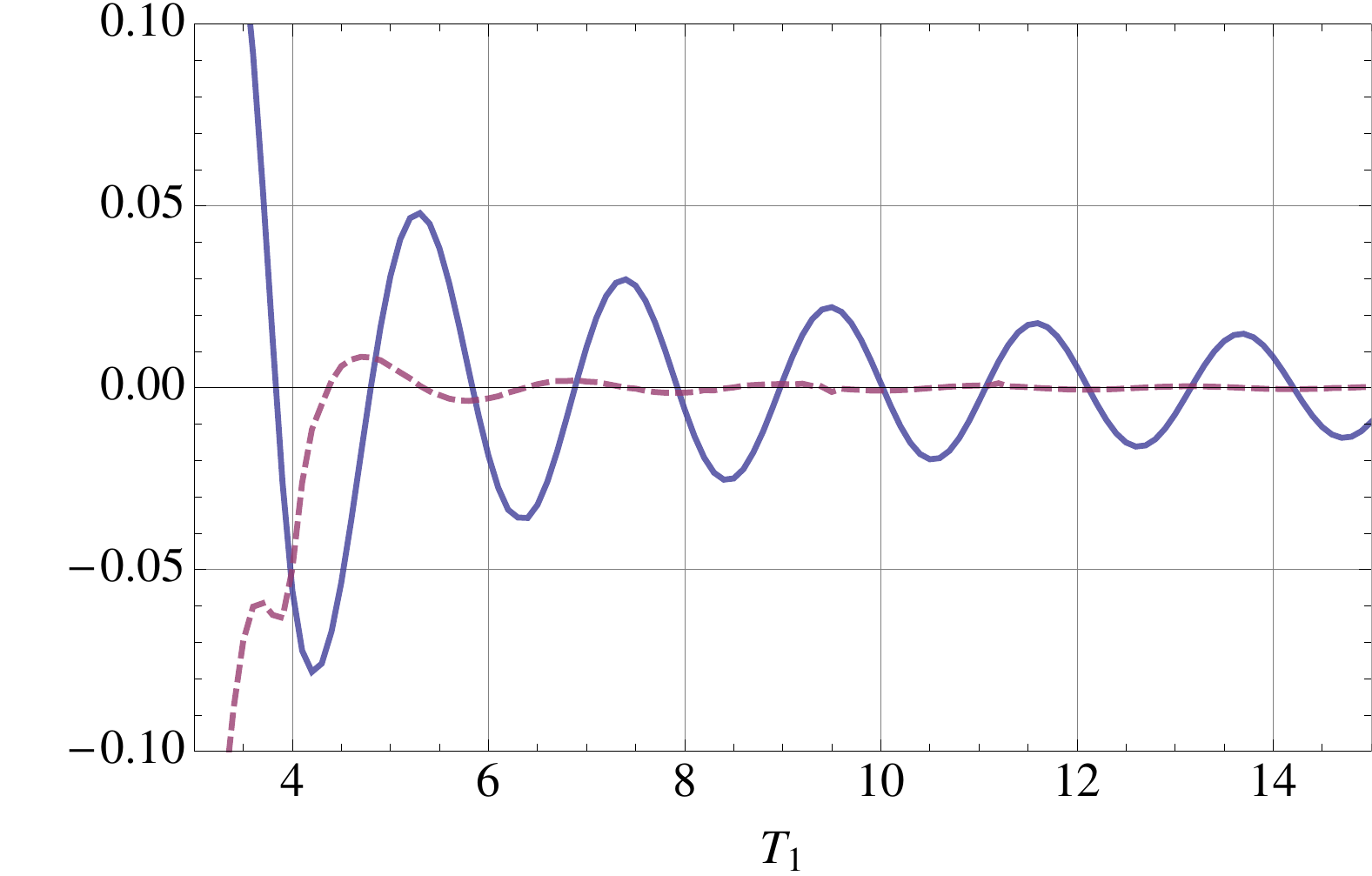}
\caption{ Signalling contributions to $\exptval{H_B}$ from \eqref{eq:signallingPe} (solid) and to $\exptval{H_f}$ from \eqref{eq:fieldHaminteract} (dashed) divided by $\lambda_A \lambda_B$ for two detectors of frequency $\Omega_A=\Omega_B=3$ at a distance $L=1$ in 2+1D Minkowski space.
Alice starts out in the state $(\ket{e_A}-\ii\ket{g_A})/\sqrt2$ and is switched on for $t=0...3$.
Bob starts in the state $ (\ket{e_B}+\ket{g_B})/\sqrt2$ and is switched on for $t=T_1...T_1+3$.
For $T_1>4$ the detectors are timelike separated. }
	\label{fig:3dim}
\end{figure}

In 1+1 dimensions we can solve \eqref{eq:bobintHinterac} analytically: 
\begin{align}\label{eq:HI11}
\exptval{\stripintHd{B}(t)}\stackrel{\mathcal{O}(\lambda_A \lambda_B)}{=}&\frac{2 \lambda_A \lambda_B}{\Omega_A} \Re\left( \alpha_B \beta_B^* e^{-\ii \Omega_B t}\right)\nonumber\\
&\quad\times\Im\left[ \alpha_A \beta_A^* \left(e^{-\ii \Omega_A T_A}-1\right)\right]
\end{align}
for $t>T_A+L$.
For timelike separated detectors, the signalling contribution to  $\exptval{H_B}$ is given by the difference between the values \eqref{eq:HI11} takes  when Bob's detector is switched on and off.
Notice that, to leading order, Bob cannot receive energy from Alice via the field because the energy of the field has no contribution proportional $\lambda_A \lambda_B$, i.e., Bob receives information from Alice without receiving energy from her.

\itsection{2+1 Dimensions}
In 2+1 dimensions there are signalling contributions also to the field energy  for timelike separated detectors. This means all three parts of the Hamiltonian, $\langle H_B \rangle$, $\langle\stripintHd{B}\rangle$ and $\langle H_f\rangle $ are affected.

From \eqref{eq:dHdt}, and the considerations thereafter, we know that the sum of the signalling contributions to $\exptval{H_B}$ and $\exptval{H_f}$ is equal to the signalling contributions to $\langle\stripintHd{B}\rangle(T_1)-\langle\stripintHd{B}\rangle(T_2)$, i.e., the difference between the work required to switch the detector on and off.

Figure \ref{fig:3dim} shows that for timelike separations the signalling contributions to  $\langle{H_f}\rangle $  are much smaller and decay faster than the ones to  $\langle H_B\rangle$. Also, all signalling contributions change sign  when either of the detectors' initial state is replaced by an orthogonal state. This indicates that also in 2+1 dimensions the timelike signalling contributions cannot be understood as absorption of energy that is sent from Alice to Bob via the field. Instead, whether Bob's coupling to the field increases or decreases the energy of his detector and the field depends on the specific properties of the state that Alice prepared for the field by coupling her detector to the field.

\itsection{Conclusions}
Intuitively, one may expect that information sent from Alice to Bob in the vacuum via a massless field necessarily travels at the speed of light. We showed that this is not the case if the strong Huygens principle is violated. This is generically the case in spacetimes of any dimension if they possess curvature. And it is the case in 1+1 and  $2n+1$D spacetime even when flat. For the latter cases, we worked out the properties of inside-the-lightcone communication explicitly. 

In particular, we found that massless inside-the-light\-cone communication does not require the transmission of energy from the sender to the receiver. Instead, the receiver requires a signal-dependent amount of work to switch his detector on and off.  In this context, recall that the  Casimir-Polder effect also sets in  when only a single particle is coupled to a field \cite{Alvaro}.  In our scenario, Alice is modulating the Casimir-Polder effect that arises from Bob's local interaction with the field. Indeed both, Casimir forces and the type of communication described in this letter arise already at second order in perturbation theory \cite{Alvaro}. This is in contrast to  communication mediated by energy-carrying quanta, which arises only from the fourth order in perturbation theory.

\itsection{Acknowledgements}
The authors would like to thank Jorma Louko for his interesting always insightful comments. We also thank William Donnelly and Ray McLenaghan for useful discussions.  The authors gratefully acknowledge support through the Ontario Trillium Scholarship and NSERC's Banting and Discovery Programs.

\bibliography{qcc_refs}

\end{document}